%% file: paper.tex
\documentclass{xsurv_article}
\usepackage{psfig}
\input {epsf_mn.tex}

\def\figinsert#1#2{\epsfbox{#1} \message{#2} }    

\begin{document}

\begin{titlepage}
\setcounter{page}{1}
\makeheadline

\title {Faint galaxies and the X-ray background}

\author{{\sc O. Almaini}, Cambridge, England \\
\medskip
{\small Institute of Astronomy} \\
\bigskip
{\sc T. Shanks, K.F. Gunn}, Durham, England\\
\medskip
{\small University of Durham} \\
\bigskip
{\sc B.J. Boyle}, Sydney, Australia\\
\medskip
{\small Anglo-Australian Observatory} \\
\bigskip
{\sc I. Georgantopoulos}, Athens, Greece\\
\medskip
{\small National Observatory of Athens} \\
\bigskip
{\sc R.E. Griffiths}, Pittsburgh, U.S.A.\\
\medskip
{\small Carnegie Mellon University} \\
\bigskip
{\sc G.C. Stewart, A.J. Blair}, Leicester, England\\
\medskip
{\small University of Leicester} \\
}

\date{Received; accepted } 
\maketitle

\summary We summarise our recent work on the faint galaxy contribution
to the cosmic X-ray background (XRB). At bright X-ray fluxes (in the
ROSAT pass band), broad line QSOs dominate the X-ray source
population, but at fainter fluxes there is evidence for a significant
contribution from emission-line galaxies. Here we present statistical
evidence that these galaxies can  account for a large fraction
of the XRB. We also demonstrate that these galaxies have significantly
harder X-ray spectra than QSOs in the ROSAT band. Finally we present
preliminary findings from infra-red spectroscopy on the nature of this
X-ray emitting galaxy population.  We conclude that a hybrid
explanation consisting of obscured/Type 2 AGN surrounded by starburst
activity can explain the properties of these galaxies and perhaps the
origin of the entire XRB.  END


\keyw
Active galactic nuclei, X-ray background, deep surveys
END

\AAAcla
END
\end{titlepage}

%

\kap{Introduction}
 
Deep ROSAT surveys have revealed that broad-line QSOs account for at
least $30 \%$ of the total $0.5-2$\,keV XRB.  Analyses of the number
count distribution and luminosity function of QSOs, however, suggests
that they are unlikely to form more than $\sim 50 \%$ of the XRB at
these energies (Boyle et al 1994, Georgantopoulos et al 1996),
although we note that recent work by Schmidt et al (1997) and Hasinger
et al (1997) disputes this claim.  Furthermore, it has been known for
many years that the X-ray spectra of QSOs are much steeper than the
spectrum of the XRB, leading to the so-called `spectral paradox'
(Gendreau et al 1995). A faint source population with a flatter X-ray
spectrum is required to account for the remainder of this background
radiation.

In recent years it has become clear that a population of X-ray
luminous narrow emission-line galaxies (NELGs) could provide a
possible explanation, perhaps dominating the X-ray source counts at
faint fluxes (Boyle et al 1995,  McHardy et
al. 1997, Carballo et al 1995, Griffiths et al 1996). We  refer
to these X-ray loud objects as `narrow-line X-ray galaxies' (NLXGs)
to distinguish them from the field galaxy population.
The nature of the X-ray emitting mechanism in these galaxies will be
discussed further in Section 4, but we will will conclude
that many
of these objects  contain low luminosity or obscured AGN (see
eg. Comastri et al 1995). In this sense the entire XRB could still be
due to `AGN' rather than two distinct classes of X-ray source.  The
most significant problem in assigning faint galaxies to X-ray sources
comes from the confusion associated with the relatively large X-ray
error box (see Hasinger et al 1997). To overcome this confusion, one
must adopt a statistical approach.

\begin{figure}
\hbox{
\psfig{figure=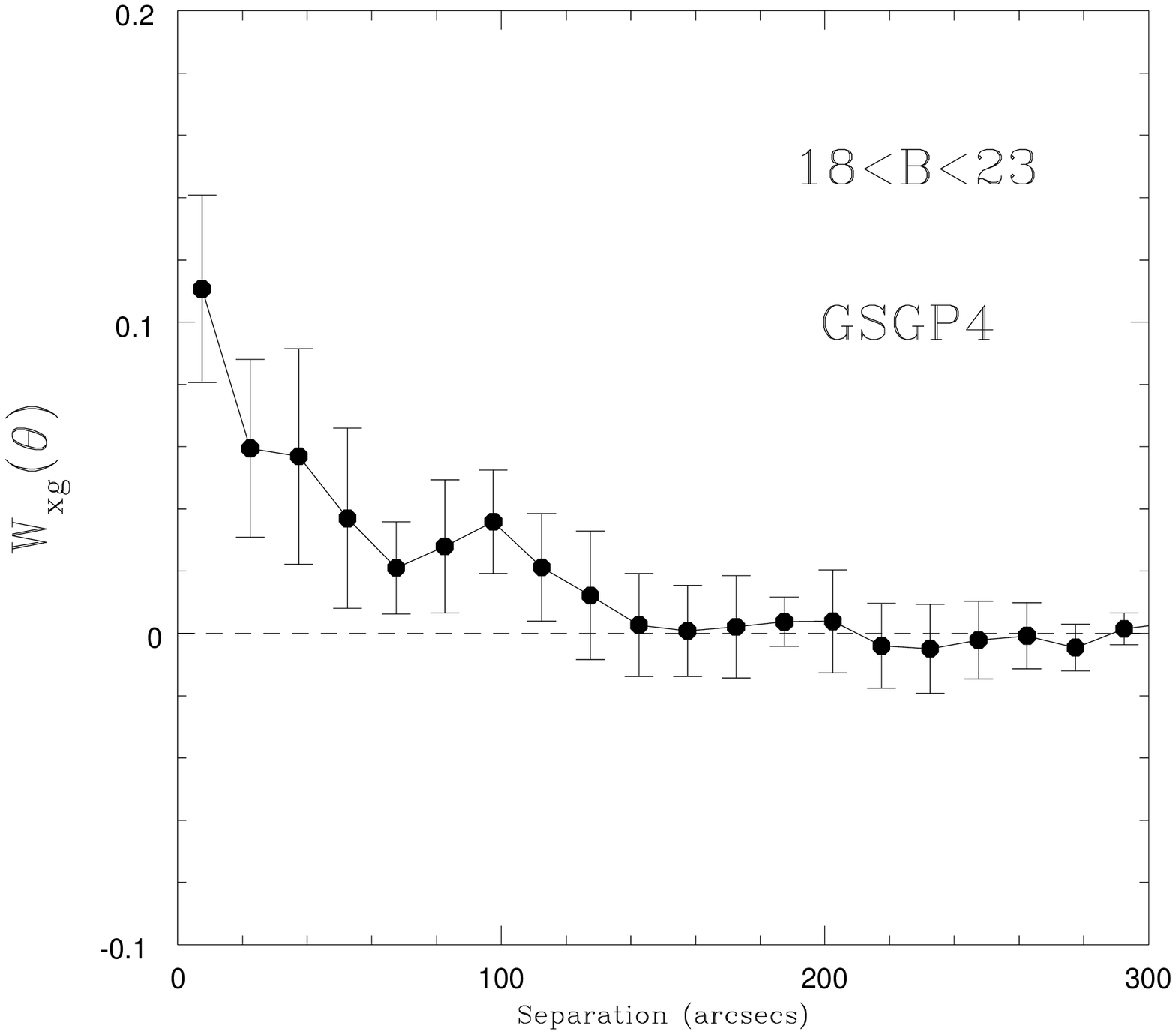,width=50mm,height=45mm}
\hspace{2mm}
\psfig{figure=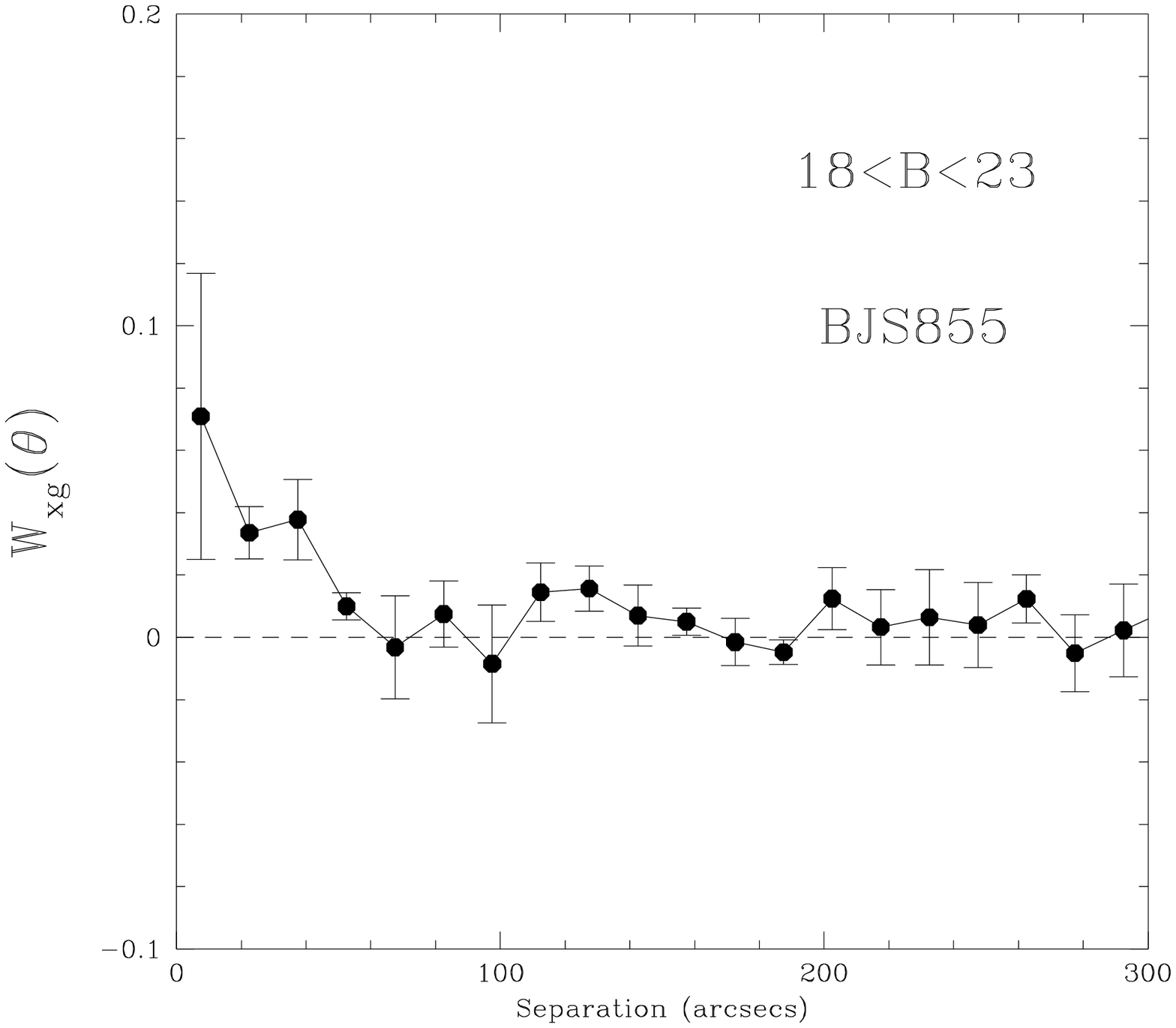,width=50mm,height=45mm}
\hspace{2mm}
\psfig{figure=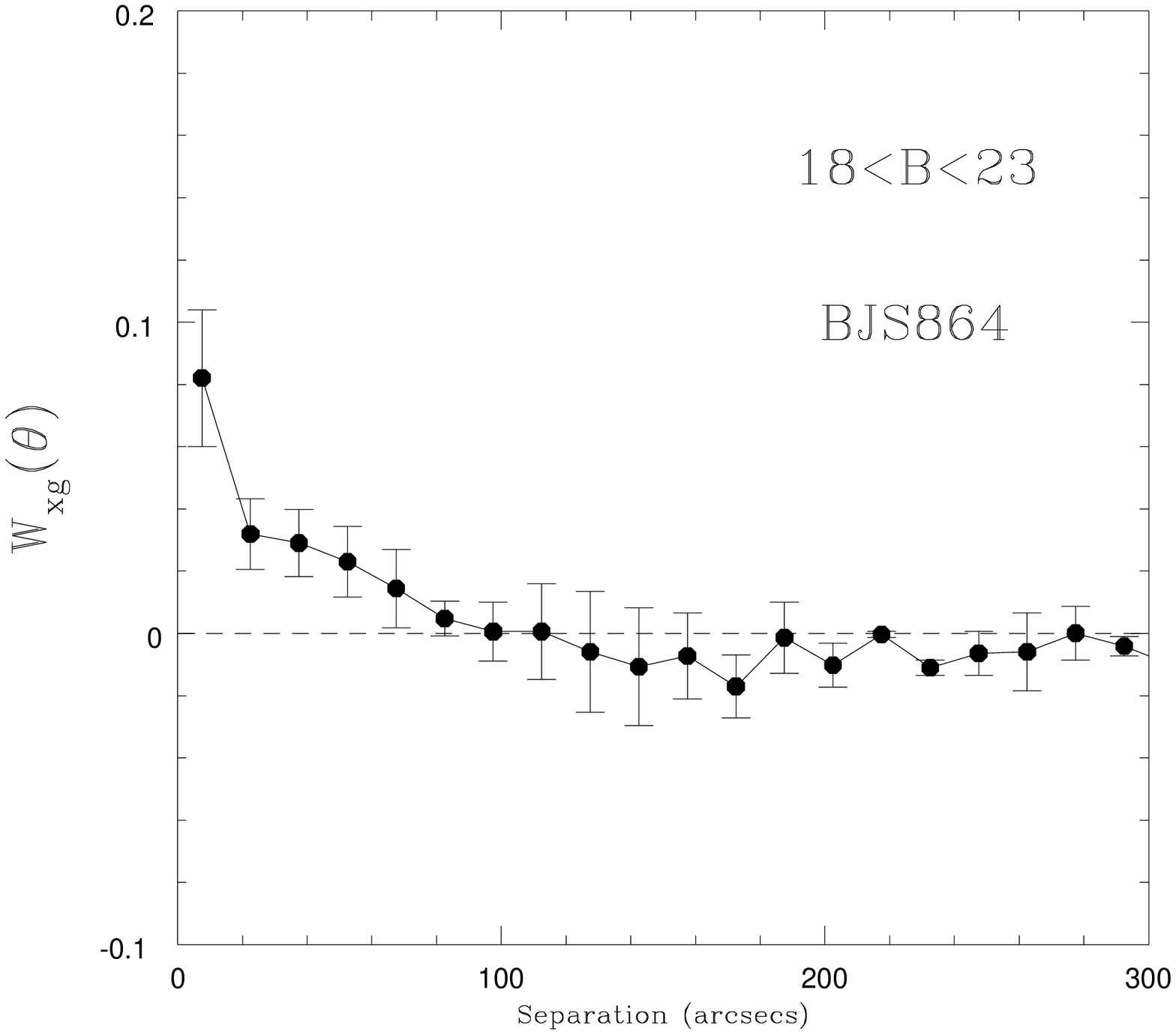,width=50mm,height=45mm}
}
\caption{The  cross-correlation function $W_{xg}(\theta)$  
of the unresolved $0.5-2$keV X-ray background with faint  $18\leq B \leq 23$ 
galaxies on 3 deep fields (see Almaini et al 1997b). }
\end{figure}

\kap{Probing the galaxy contribution by cross-correlation techniques}

The clearest evidence that faint galaxies could be important
contributors to the XRB came from the study of Roche et al (1995), who
cross-correlated faint ($B<23$) galaxy catalogues with deep X-ray
observations, thus avoiding the problems associated with source
confusion. In Almaini et al (1997b) we performed an independent test
of the Roche et al results on new deep ($\sim$50ks) $\it ROSAT$
exposures and for the first time attempted to measure the evolution in
the X-ray emissivity of faint galaxies with redshift.  The
cross-correlation of unidentified X-ray sources with faint galaxies
indicated that these could account for $ 20 \pm 7 \%$ of all X-ray
sources to a limiting flux of
$\sim4\times10^{-15}$erg$\,$s$^{-1}$cm$^{-2}$ in the $0.5-2.0$\,keV
band. To probe deeper, cross-correlations were also performed with the
residual, unresolved XRB images. Significant signals were again
obtained on all 3 $\it ROSAT$ fields (see Figure 1), independently
confirming the results obtained by Roche et al (1995).

To constrain the evolution of X-ray emissivity with redshift, we used
the magnitudes of the thousands of catalogue galaxies as probes of
their redshift distribution, adapting the formalism of Treyer \& Lahav
(1996).  We found evidence for very strong evolution in the mean X-ray
luminosity of the form $\left\langle{L_x}\right\rangle \propto
(1+z)^{3.22\pm 0.98}$. This represents the first evidence that the
X-ray emission from faint galaxies evolves as strongly as AGN. Similar
results were obtained by analysing small numbers of bright narrow
emission-line galaxies emerging from deep $\it ROSAT$ exposures
(Griffiths et al 1996, Boyle et al 1995). A simple extrapolation to
z=2 suggests that faint galaxies can account for $\sim 40 \pm 10 \%$
of the total XRB at 1keV.

\begin{figure}
\hbox{
\psfig{figure=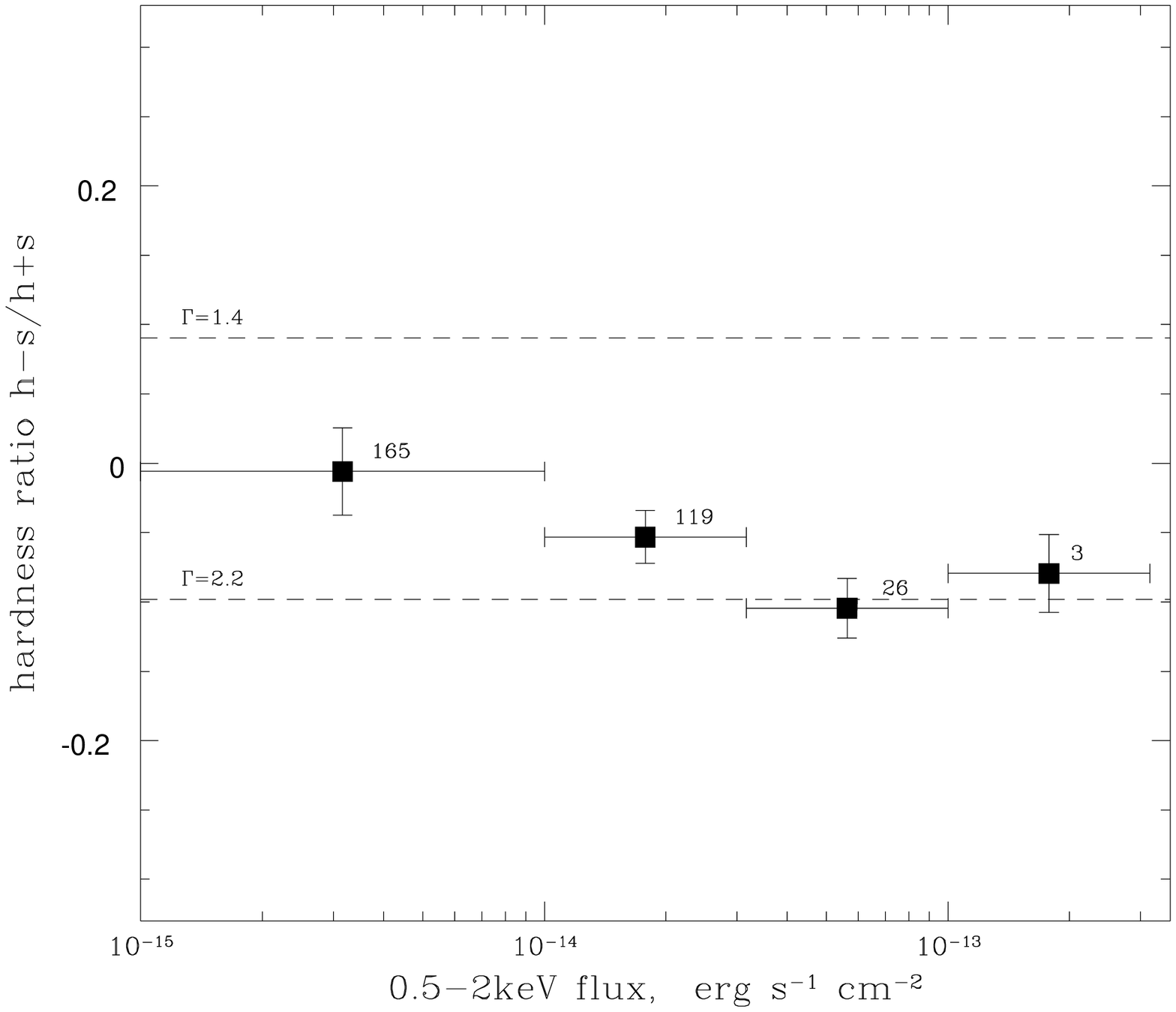,width=85mm,height=65mm}
\hspace{2mm}
\psfig{figure=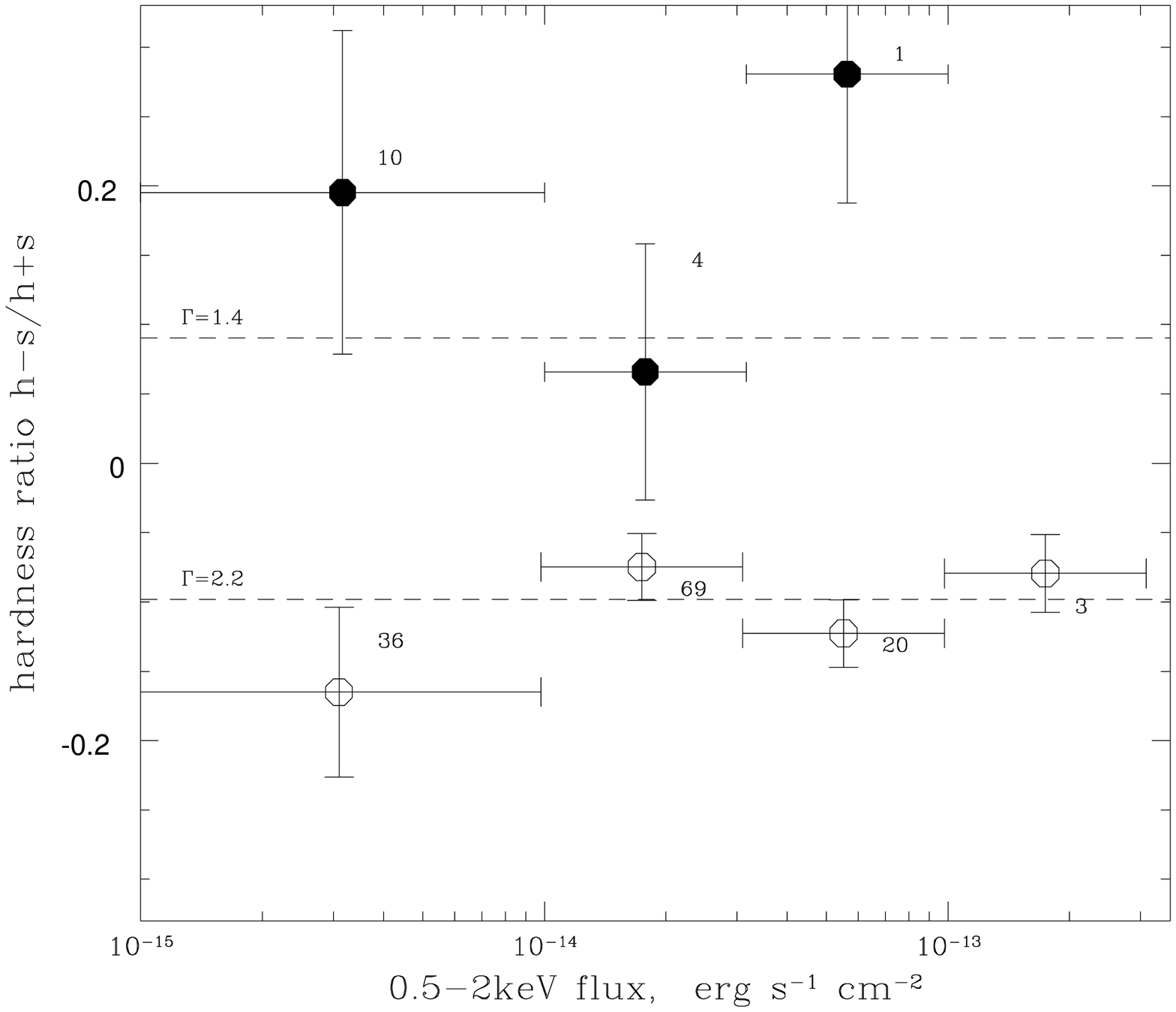,width=85mm,height=65mm}
}
\caption{(a) 0.5-2\,keV hardness ratios as a function of flux for the
stacked spectra of all X-ray sources from 5 deep ROSAT fields.  In (b)
we separate the QSOs (unfilled circles) from the subset of 15 narrow
emission line galaxies considered to be the most unambiguous
identifications (filled circles).  The number of X-ray sources within
each bin is indicated. For comparison, the hardness ratios for two
power-law models are also shown.}
\end{figure}

\kap{The X-ray spectra of faint ROSAT sources}

In Almaini et al (1996) we presented an analysis of the X-ray spectra
of all X-ray sources from our deep survey. The results confirmed the
trend found by Hasinger et al (1993) and Vikhlinin et al (1995) that
the average source spectra harden towards fainter fluxes.  Hardness
ratios for the stacked spectra of all sources as a function of flux
are shown on Figure 2(a).

Using the optical identifications available for most of these sources,
it was found that ordinary, broad-line QSOs are not responsible for
this spectral hardening (Figure 2b). This implied that the change in
mean spectra was due to the emergence of another source
population. Our cross-correlation results suggest that $20\pm 7$\% of
the remaining, unidentified sources are due to faint galaxies, but
confusion problems prevent us from determining exactly which X-ray
sources are responsible.  We therefore selected a restricted sample of
the most likely galaxy candidates with brighter optical magnitudes
($B<21.5$) and lying within 20 arcsec of the X-ray source. The
hardness ratios for the 15 emission-line galaxies satisfying this
criteria are shown in Figure 2b. We expect approximately 3 of these
galaxies to be spurious identifications, but nevertheless it is clear
that their mean X-ray spectra are significantly harder than those of
QSOs (Almaini et al 1996, Romero-Colmenero et al 1996). Hence the
X-ray spectra provide further evidence that NLXGs may finally provide
the solution to the origin of the XRB. The key test is to examine
their mean spectral properties with a large sample at higher energies,
which will be possible with AXAF or XMM.

\begin{figure}[t]
\centering \centerline{\epsfxsize=9truecm \figinsert{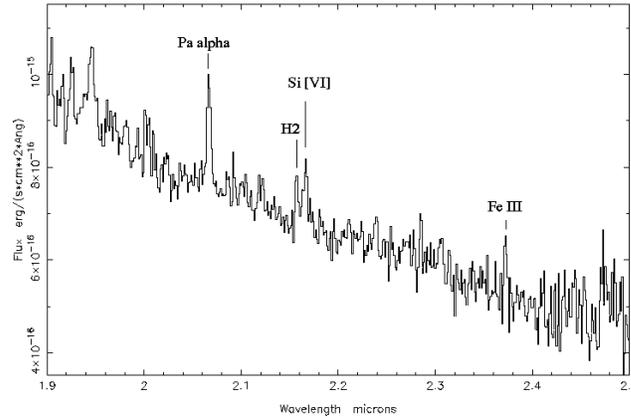}{0.0pt}}
\caption{Infra-red spectrum of a NLXG at redshift
$z=0.105$.  The coronal Si[VI] line proves the existence of an AGN,
although the non-detection of broad Pa$\alpha$ implies an obscuring column
$n_H > 5\times 10^{22}$atom cm$^{-2}$.}
\end{figure}

\kap{Understanding the nature of NLXGs}

There is mounting evidence that a population of X-ray luminous
`galaxies' could finally provide a solution to the origin of the XRB.
The important question now is to understand the nature of this unusual
activity. Using standard optical emission line ratios, these objects
appear on the borderline between starburst and Seyfert 2
classification and in most cases the distinction is very ambiguous
(Boyle et al 1995, McHardy et al 1997). One possibility is that
these are a new type of evolved starburst galaxy in which the X-ray
emission comes from massive X-ray binaries (Griffiths and Padovani
1990). No local starburst galaxies have been found with such hard
X-ray spectra however.  These galaxies are also significantly more
X-ray luminous than any known starburst galaxies, and their far
infra-red fluxes are generally too low for the X-ray emission to be
entirely due to starforming activity (Iwasawa et al 1997).  The most
likely explanation is that the hard X-rays come from an obscured
active nucleus. Such models can readily reproduce the spectrum of the
X-ray background (Comastri et al 1995). Recent work by Hasinger et al
(1997; see also these proceedings) has  suggested that many of
these galaxies would indeed be classified as AGN in higher signal to
noise optical spectra, albeit as low luminosity and/or type 2 AGN.

If these objects are obscured AGN, one might expect to detect broad
emission lines in the infra-red waveband (the dust extinction at $K$
is approximately 6 magnitudes lower than in the $V$ band). We have
recently obtained UK Infra-Red Telescope CGS4 spectroscopy for a small
number of NLXGs. Of the 5 objects studied so far, none show any
evidence for the expected broad Paschen emission lines.  In most
cases, this implies an absorbing column of at least $5\times 10^{22}$
atom cm$^{-2}$ (if an AGN is present). Such columns are inconsistent
with the large ROSAT luminosities below 1keV, unless there is some
contribution from an additional source of soft X-ray flux. In at
least one object, the detection of highly ionized Si[VI] clearly
indicates the presence of an AGN, despite the non-detection of broad
Paschen lines (Figure 3). As such, this is similar to the optical
narrow-line objects detected by Hasinger et al (1997). A plausible way
to account for these properties is to postulate a hybrid model,
consisting of an obscured AGN surrounded by starforming activity.
Such a model would certainly explain the ambiguous optical line ratios
seen in these objects. The hard ASCA NLXG detected by Iwasawa et al
(1997) shows clear evidence for this dual behaviour. As discussed in
Fabian et al (1997), obscuration by nuclear starburst activity may be
an inevitable consequence of triggering a low luminosity AGN, and
could perhaps account for the obscuration of the entire hard XRB.

\kap{Conclusions}

There is now overwhelming evidence for the existence of a population
of X-ray luminous galaxies, which could explain the remainder of the
XRB. Our cross-correlation analysis suggests that these objects could
account for $40 \pm 10\% $ of the XRB at 1keV. In addition, the
individually identified galaxies  show
significantly harder X-ray spectra than QSOs, more consistent with
that of the XRB. Obscured and/or low luminosity AGN provide the most
natural explanation for this activity, although our infra-red
spectroscopy and the optical properties of these galaxies suggest
the presence of an additional soft X-ray component. A hybrid model
consisting of an obscured AGN surrounded by starburst activity
provides the most natural explanation for all observed properties.

\sect{References}

\rf{Almaini O., Shanks T., Boyle B.J., Griffiths R.E., Roche N.,  Stewart G.C.
\& Georgantopoulos I.,  1996, MNRAS 282, 295}
\rf{Almaini O. \& Fabian A.C., 1997a, MNRAS 288, L19}
\rf{Almaini O., Shanks T.,  Griffiths R.E., Boyle B.J., Roche N.,  
Georgantopoulos I., \& Stewart G.C.   1997b, MNRAS 291, 372}
\rf{Boyle B.J., Griffiths R.E., Shanks T., Stewart G.C.,
Georgantopoulos I., 1994, MNRAS, 271,639}
\rf{Boyle B.J., McMahon R.G., Wilkes B.J., \& Elvis M.,
 1995, MNRAS 276, 315}
\rf{Carballo R. et al 1995, MNRAS 277, 1312}
\rf{Comastri A., Setti G., Zamorani G. \& Hasinger G.,
1995,  A\&A, 296, 1}
\rf{Fabian A.C., Barcons X., Almaini O., 1997, MNRAS submitted}
\rf{Gendreau K.C. et al, 1995, Publ. Astron. Soc. Japan, 47, L5-L9}
\rf{Georgantopoulos I., Stewart G.C., Shanks T., Griffiths R.E., \& Boyle B.J.,
1996, MNRAS 280, 276}
\rf{Griffiths R.E. \& Padovani P. 1990, ApJ 360, 483}
\rf{Griffiths R.E., Della Ceca R., Georgantopoulos I., Boyle B.J., 
Stewart G.C., \& Shanks T., 1996, MNRAS 281, 71}
\rf{Hasinger G., Burg R., Giacconi R., Hartner G., Schmidt M.,
Tr\"{u}mper J., Zamorani G., 1993, A\&A, 275, 1}
\rf{Hasinger G., Burg R., Giacconi R., Schmidt M., Trumper J. \& Zamorani G., 
 1997, A\&A in press}
\rf{Iwasawa K., Fabian A.C., Brandt W.N., Crawford C.S.,
Almaini O., 1997, MNRAS, 291, L17}
\rf{McHardy I. et al, 1997, MNRAS, In press}
\rf{Roche N., Shanks T., Georgantopoulos I., Stewart G.C., Boyle B.J., \& 
Griffiths R.E., 1995, MNRAS 273, L15}
\rf{Schmidt M. et al, 1997, A\&A in press}
\rf{Romero-Colmenero E. et al,  1996, MNRAS 282, 94}
\rf{Shanks T., Georgantopoulos I.,  Stewart G.C., Pounds K.A., Boyle B.J. 
\& Griffiths R.E., 1991 Nat 353, 315}
\rf{Treyer M.A. \& Lahav O. 1996 MNRAS 280, 469}
\rf{Vikhlinin A., Forman W., Jones C. \&  Murray S., 
1995 ApJ 451, 564}

%

\address 
\rf{Omar Almaini, Institute of Astronomy, Madingley Road, Cambridge, CB3 OHA, e-mail: omar@ast.cam.ac.uk} 
END

\end{document}

%% file: epsf_mn.tex
\newread\epsffilein    
\newif\ifepsffileok    
\newif\ifepsfbbfound   
\newif\ifepsfverbose   
\newdimen\epsfxsize    
\newdimen\epsfysize    
\newdimen\epsftsize    
\newdimen\epsfrsize    
\newdimen\epsftmp      
\newdimen\pspoints     
\def\epsfbox#1{\global\def\epsfllx{72}\global\def\epsflly{72}%
   \global\def\epsfurx{540}\global\def\epsfury{720}%
   \def\lbracket{[}\def\testit{#1}\ifx\testit\lbracket
   \let\next=\epsfgetlitbb\else\let\next=\epsfnormal\fi\next{#1}}%
\def\epsfgetlitbb#1#2 #3 #4 #5]#6{\epsfgrab #2 #3 #4 #5 .\\%
   \epsfsetgraph{#6}}%
\def\epsfnormal#1{\epsfgetbb{#1}\epsfsetgraph{#1}}%
\def\epsfgetbb#1{%
\openin\epsffilein=#1
\ifeof\epsffilein\errmessage{I couldn't open #1, will ignore it}\else
   {\epsffileoktrue \chardef\other=12
    \def\do##1{\catcode`##1=\other}\dospecials \catcode`\ =10
    \loop
       \read\epsffilein to \epsffileline
       \ifeof\epsffilein\epsffileokfalse\else
          \expandafter\epsfaux\epsffileline:. \\%
       \fi
   \ifepsffileok\repeat
   \ifepsfbbfound\else
    \ifepsfverbose\message{No bounding box comment in #1; using defaults}\fi\fi
   }\closein\epsffilein\fi}%
\def\epsfsetgraph#1{%
   \epsfrsize=\epsfury\pspoints
   \advance\epsfrsize by-\epsflly\pspoints
   \epsftsize=\epsfurx\pspoints
   \advance\epsftsize by-\epsfllx\pspoints
   \epsfxsize\epsfsize\epsftsize\epsfrsize
   \ifnum\epsfxsize=0 \ifnum\epsfysize=0
      \epsfxsize=\epsftsize \epsfysize=\epsfrsize
     \else\epsftmp=\epsftsize \divide\epsftmp\epsfrsize
       \epsfxsize=\epsfysize \multiply\epsfxsize\epsftmp
       \multiply\epsftmp\epsfrsize \advance\epsftsize-\epsftmp
       \epsftmp=\epsfysize
       \loop \advance\epsftsize\epsftsize \divide\epsftmp 2
       \ifnum\epsftmp>0
          \ifnum\epsftsize<\epsfrsize\else
             \advance\epsftsize-\epsfrsize \advance\epsfxsize\epsftmp \fi
       \repeat
     \fi
   \else\epsftmp=\epsfrsize \divide\epsftmp\epsftsize
     \epsfysize=\epsfxsize \multiply\epsfysize\epsftmp   
     \multiply\epsftmp\epsftsize \advance\epsfrsize-\epsftmp
     \epsftmp=\epsfxsize
     \loop \advance\epsfrsize\epsfrsize \divide\epsftmp 2
     \ifnum\epsftmp>0
        \ifnum\epsfrsize<\epsftsize\else
           \advance\epsfrsize-\epsftsize \advance\epsfysize\epsftmp \fi
     \repeat     
   \fi
   \ifepsfverbose\message{#1: width=\the\epsfxsize, height=\the\epsfysize}\fi
   \epsftmp=10\epsfxsize \divide\epsftmp\pspoints
   \newcount\figskipcount
      \message{#1 \the\epsfysize  }
   \vbox to\epsfysize{\vfil\hbox to\epsfxsize{%
      \includegraphics{#1}%
      \hfil}}%
\epsfxsize=0pt\epsfysize=0pt}%

{\catcode`\%=12 \global\let\epsfpercent=
\long\def\epsfaux#1#2:#3\\{\ifx#1\epsfpercent
   \def\testit{#2}\ifx\testit\epsfbblit
      \epsfgrab #3 . . . \\%
      \epsffileokfalse
      \global\epsfbbfoundtrue
   \fi\else\ifx#1\par\else\epsffileokfalse\fi\fi}%
\def\epsfgrab #1 #2 #3 #4 #5\\{%
   \global\def\epsfllx{#1}\ifx\epsfllx\empty
      \epsfgrab #2 #3 #4 #5 .\\\else
   \global\def\epsflly{#2}%
   \global\def\epsfurx{#3}\global\def\epsfury{#4}\fi}%
\def\epsfsize#1#2{\epsfxsize}